\documentclass[12pt,manuscript]{aastex}
\usepackage{soul}

\slugcomment{}

\shorttitle{}
\shortauthors{}

\begin{document}

\title{\418 - HOW DOES A YOUNG NEUTRON STAR SPIN DOWN TO A 9 s PERIOD WITH A DIPOLE FIELD LESS~THAN 10$^{13}$ G ?}

\author{M.A. Alpar,  \"{U}. Ertan \&  \c{S}. \c{C}al{\i}\c{s}kan}
\affil{Sabanc\i\ University, 34956, Orhanl\i\, Tuzla, \.Istanbul,
Turkey}

\email{unal@sabanciuniv.edu}

\def\be{\begin{equation}}
\def\ee{\end{equation}}
\def\ba{\begin{eqnarray}}
\def\ea{\end{eqnarray}}
\def\m{\mathrm}
\def\a{\alpha}
\def\Mdot*{\dot{M}_*}
\def\Mdotin{\dot{M}_{\mathrm{in}}}
\def\Mdot{\dot{M}}
\def\Pdot{\dot{P}}
\def\Msun{M_{\odot}}
\def\Lin{L_{\mathrm{in}}}
\def\Rin{R_{\mathrm{in}}}
\def\rin{r_{\mathrm{in}}}
\def\rlc{r_{\mathrm{LC}}}
\def\rout{r_{\mathrm{out}}}
\def\rco{r_{\mathrm{co}}}
\def\Rout{R_{\mathrm{out}}}
\def\Ldisk{L_{\mathrm{disk}}}
\def\Lx{L_{\mathrm{x}}}
\def\Md{M_{\mathrm{d}}}
\def\cs{c_{\mathrm{s}}}
\def\dEb{\delta E_{\mathrm{burst}}}
\def\dEx{\delta E_{\mathrm{x}}}
\def\Bstar{B_\ast}
\def\Bb{\beta_{\mathrm{b}}}
\def\Be{\beta_{\mathrm{e}}}
\def\Rc{\R_{\mathrm{c}}}
\def\rA{r_{\mathrm{A}}}
\def\rp{r_{\mathrm{p}}}
\def\Tp{T_{\mathrm{p}}}
\def\dMin{\delta M_{\mathrm{in}}}
\def\dM*{\delta M_*}
\def\Teff{T_{\mathrm{eff}}}
\def\Tirr{T_{\mathrm{irr}}}
\def\Firr{F_{\mathrm{irr}}}
\def\Tcrit{T_{\mathrm{crit}}}
\def\P0min{P_{0,{\mathrm{min}}}}
\def\Av{A_{\mathrm{V}}}
\def\ah{\alpha_{\mathrm{hot}}}
\def\ac{\alpha_{\mathrm{cold}}}
\def\p{\propto}
\def\m{\mathrm}
\def\fast{\omega_{\ast}}
\def\Alfven{Alfv$\acute{e}$n}
\def\418{SGR 0418+5729}
\def\142{AXP 0142+61}
\def\2144{PSR J2144-3933}
\def\Mdotacc{\dot{M}_{\mathrm{acc}}}

\begin{abstract}
The period derivative bound for SGR 0418+5729 (Rea et al. 2010) establishes the
magnetic dipole moment to be distinctly lower than the magnetar
range, placing the source beyond the regime of isolated pulsar
activity in the $P-\dot{P}$ diagram and giving a characteristic 
age $>$ 2  $\times$ 10$^{7}$ years, much older than the 10$^{5}$ year age 
range of SGRs and AXPs. So the spindown must be produced by a 
mechanism other than dipole radiation in vacuum. A fallback disk 
will spin
down a neutron star with surface dipole magnetic field in the
10$^{12}$ G range and initial rotation period $P_0 \sim 100$ ms to
the 9.1 s period of SGR 0418+5729 in a few $10^4$ to $\sim 10^5$ years. The
current upper limit to the period derivative gives a lower limit
of $\sim 10^5$ years to the age that is not sensitive to the
neutron star's initial conditions. The  total 
magnetic field on the surface of SGR 0418+5729 could be 
significantly larger than its 10$^{12}$ G dipole component.
\end{abstract}


\section{INTRODUCTION}

The recently discovered \418 (van der Horst et al. 2010) has a
period $P=9.1$ s (G\"{o}\u{g}\"{u}\c{s}, Woods \& Kouveliotou
2009) in the narrow range of AXP and SGR periods (Mereghetti
2008). The spin-down rate has not been measured yet (Kuiper \&
Hermsen 2009, Woods, G\"{o}\u{g}\"{u}\c{s} \& Kouveliotou 2009,
Esposito et al. 2010, Rea et al. 2010). The  best period derivative upper
limit, $\dot{P} < 6 \times 10^{-15}$ s s$^{-1}$ (Rea et al. 2010),
evaluated as dipole spin-down of an isolated star, gives a
surface dipole magnetic field $B_0 < 1.5 \times 10^{13}$ G at the
poles, much lower than fields previously deduced from spin-down
rates of magnetars. The characteristic age $P/(2\dot{P}) > 2.5
\times 10^7$ yrs, while AXPs and SGRs, some of which are
associated with SNR (Esposito et al. 2009, Mereghetti 2008 and
references therein) are believed to be young neutron stars with
ages $\sim 10^5$ yr. \418 is similar to other AXPs and SGRs in all
observed properties except for $\dot{P}$. The energy in its soft
gamma-ray bursts requires a total magnetic field $\sim 10^{12}$ G
on the neutron star surface, but if all SGRs have super-outbursts
occasionally, so far observed from SGR 1806-20, SGR 0526-66 and
SGR 1900+14, the total surface magnetic field $B_{\m{total}} \sim
10^{14} - 10^{15}$ G according to the magnetar model (Duncan \&
Thompson 1992; Thompson \& Duncan 1995). If \418 is a standard
magnetar, it provides a clear counterexample to the proposition
that for magnetars the dipole component of the magnetic field is
of the same order as the total field.

If the spindown to the present period was achieved by
magnetic dipole radiation \418 would be an exceptional object,
mimicking all SGR and AXP properties while not belonging to the
class. Its position in the $P-\dot{P}$ diagram, beyond the so
called death-valley, makes it exceptional also among the rotation
powered isolated pulsars: the only other source located similarly
in $P-\dot{P}$ is the radio pulsar \2144.
Furthermore, if \418 is older than $2.5 \times 10^7$ yrs as its
characteristic age $P/(2\dot{P})$ suggests, its quiescent X-ray
luminosity cannot be explained by cooling, reheating or magnetic
field decay, let alone explaining soft gamma-ray outbursts
occurring at such old age. If \418 is much younger than its
characteristic age with dipole spin-down, its initial rotation
period would have to be close to the present 9.1 s period, again
making this source unique, standing far out from the initial
period distribution inferred from population synthesis
(Faucher-Gigu\'ere \& Kaspi 2006).

The dipole component of the field $B_0$ determines torques
due to electromagnetic radiation and interactions with the
environment. Estimates of $B_0$ from spin-down rates
depend on the torque mechanism. The total surface magnetic field
is derived from measurements of cyclotron lines (Ibrahim et al.
2002) and the spectral continuum (G\"uver et al. 2007; 2008).
Historically, the dipole field measurements came first
(Kouveliotou et al. 1998). The field inferred with the dipole
spin-down torque was in the magnetar range, supporting the
magnetar model which had been proposed to explain the SGR bursts
and other SGR and AXP properties including spindown to long
periods at a young age (Duncan \& Thompson 1992; Thompson \&
Duncan 1995). The identification of the dipole component with the
total field has been taken for granted.

We proceed, by Occam's razor, to posit that \418 is a member
of the same class of young neutron stars as the other SGRs and
AXPs, but its spin down is not due to magnetic dipole radiation.
So there must be matter around the star, in a bound state,
therefore carrying angular momentum.
For {\em isolated} neutron stars a fallback disk, which can be
formed in some supernovae (Michel 1988, Chevalier
1989, Lin et al. 1991) will provide this. The fallback disk model
was proposed by Chatterjee et al. (2000) for AXPs, and
independently by Alpar (2001) as a possible way of explaining the
different classes of young neutron stars, including the X-ray dim
isolated neutron stars (XDINs) and compact central objects (CCOs)
as well as AXPs and SGRs. The prime motivation was to address the
period clustering which strongly suggests a regulating store of
angular momentum. For a given value of $\Pdot$ the dipole moment
inferred with fallback disk model is generally less than that
derived assuming isolated dipole spin-down. The differences in
$\Pdot$ between sources of similar period are not primarily due to
differences in magnetic dipole moment, with the fallback disk also
playing a critical role in evolution. The model indicates surface
dipole fields $\sim 10^{12} - 10^{13}$ G. The bursts may be
powered by strong total magnetic fields $B_{\m{total}} \sim
10^{14} - 10^{15}$ G as in the magnetar model- implying that the
dipole field is smaller than the total field.
The discovery of a disk around \142 (Wang, Chakrabarty \&
Kaplan 2006) gave strong support to the fallback disk model. Ertan
et al (2007) showed that the entire non-pulsed optical to mid-IR
spectrum can be understood as emission from a gaseous disk, while
the pulsed optical signal is produced in the magnetosphere (Ertan \&
Cheng 2004, Cheng \& Ruderman 1991).

This Letter investigates evolutionary scenarios for \418
employing a fallback disk. We show that the period derivative as
well as the period and X-ray luminosity in quiescence are explained
quite naturally, and a fallback disk can spin down the neutron star
to a period of 9.1 s in a few 10$^4$ to $\sim 10^5$ years.
%
%
%

\section{EVOLUTION WITH A FALLBACK DISK}

The mass and mass inflow rate of the fallback disk decay through
viscous dynamics,
modified by irradiation from the neutron star. The fallback disk,
though truncated at the inner radius, follows the self-similar
solutions with power law decay in time (Pringle 1974) quite closely
as long as the entire disk is viscous (Ertan et al. 2009). Viscous
activity stops when the local temperature falls below a critical
temperature $T_p \sim$ $100$ K, becoming too cold for
sufficient ionization for the magneto-rotational instability to
generate viscosity and sustain mass inflow (Inutsuka \& Sano 2005).
Such passive regions grow starting from the outer disk. Irradiation
by the star can keep the outer disk at temperatures higher than
$T_p$ for a while, delaying the passive phase, keeping a
larger part of the disk active. This interaction between the gradual
transition to a final passive phase, and the effect of irradiation
to prolong the active phase determines the evolution in a
complicated way. To calculate the irradiation flux impinging on the
disk we employ the same irradiation efficiency as in our best fits
for the disk observed around \142 (Ertan et al. 2007, see Ertan \&
\c{C}al{\i}\c{s}kan 2006 for the other AXPs).

At each step in the evolution a solution for the entire disk is
constructed taking all these effects into account. The mass inflow
rate $\Mdotin$ arriving at the inner disk is obtained
and the inner disk radius $\rin$ is determined as
the \Alfven~ radius
\begin{equation}
\rA = 10^9~ \m{cm} ~\mu_{30}^{4/7}
(\dot{M}_{\mathrm{in}15})^{-2/7} (M/M_\odot)^{-1/7} .
\end{equation}
Here $\Msun$ is the solar mass, 
$\dot{M}_{\mathrm{in}15}$ the mass
inflow rate in 10$^{15}$ gm s$^{-1}$, and $\mu_{30}$ the dipole
magnetic moment in $10^{30}$ G cm$^3$. The important distance
scales are the light cylinder radius $\rlc = c/\Omega$, the
co-rotation radius $\rco = (GM)^{1/3}/\Omega^{2/3}$ and $\rA$. The
fallback disk will effect the evolution when the disk's inner
radius is within the neutron star's light cylinder. The effect of
the disk will decrease drastically when the disk moves outside the
light cylinder.

Throughout the evolution $\rA > \rco$, so the neutron star is
a fast rotator, and the torque applied by the disk is always a
spin-down torque. The neutron star is in the propeller regime
(Illarionov \& Sunyaev 1975). In contrast to the
original propeller picture the fallback disk model takes some
portion $\Mdotacc$ of the mass inflow $\Mdotin$ to be accreting
onto the neutron star during spin-down (Chatterjee, Hernquist \&
Narayan 2000, Alpar 2001). Rappaport, Fregeau \& Spruit (2004)
have shown from general considerations of accreting neutron stars
that partial accretion must be taking place. This provides the
X-ray luminosity in the fallback disk model throughout most of the
evolution, when $\rco < \rA < \rlc$. The luminosity evolution is
determined by the unknown fraction $\Mdotacc/\Mdotin$ and the
initial fallback disk mass $\Md$ which effects the evolution of
$\Mdotin$.

The spin-down rate of a neutron star under disk torques is given
by
\begin{equation}
I \dot{\Omega} = \Mdotin(G M \rA)^{1/2} F(\omega) \label{torque}
\end{equation}
where $I$ is the moment of inertia, $\dot{\Omega}$ the spin-down
rate, $\Omega$ the rotation rate, $\Mdotin$ the mass inflow
rate arriving from the disk at its inner boundary, and  $M$ is the
star's mass. $F(\omega)$ is the dimensionless torque which
depends on the fastness parameter $\omega \equiv
\Omega/\Omega_K(r_A)$, $\Omega_K(\rA)$ being the Keplerian
rotation rate at $\rA$.
%
A dimensionless disk torque
\begin{equation}
F(\omega) = ( 1 - \omega^2 )\cong
 -\omega^2
\end{equation}
is indicated by our earlier results (Ertan et al. 2009, Ertan \& Erkut
2008). This torque is due to the azimuthal bending of
magnetic field lines from the co-rotating magnetosphere at $\rco$
to the slower rotating inner disk at $\rA > \rco$. Equations
(1)-(3) show that the torque is independent of $\Mdotin$.
($F(\omega) \cong -\omega^{2+\delta}$ gives a weak dependence
$\propto \Mdotin^{-3\delta/7}$). We integrate $\dot{\Omega}$ to
get $\Omega$, reconstruct the disk, with current $\rA$ and $\rlc$,
irradiated by the current luminosity, and proceed by iteration.

As $\Mdotin$ decreases and the star spins down, $\rA$ increases
with time faster than $\rlc$ does.
Near and beyond the light cylinder $\rlc$ the electromagnetic field
gradually changes from the dipole magnetic field to wave fields. The
inner disk radius is somewhat larger than $\rA$ in this region.
Ek\c{s}i \& Alpar (2005) have studied the transition towards the
wave zone. They show that the disk is stable beyond the light
cylinder as long as the inner disk radius $r_{in}$ remains within a
critical distance which depends on the angle between the rotation
and magnetic axes of the star, ranging from 2.5 $\rlc$ for a
perpendicular rotator to many $\rlc$ for an almost aligned rotator.
The torque and luminosity should drop within a narrow range of
$r_{in} \cong \rlc$ -- the disk can be stable far beyond $\rlc$, but
is causally disconnected from the star and magnetosphere. Cooling or
energy dissipation in the neutron star account for a much reduced
X-ray luminosity. For the torque we consider two distinct models:
(i) We assume that the disk remains undetached from the light
cylinder and set $r_{in} = \rlc$. This can be qualitatively
justified as mass lost by the disk cannot penetrate into the
magnetosphere, but will tend to pile up around the light cylinder.
As the disk inner radius reaches $\rlc$ from inside the mass
pile-up is likely to keep $r_{in}$ from detaching from $\rlc$.
(ii) The minimal torque is the dipole radiation torque taking over
immediately when $r_{in} \geq \rlc$. The actual torque should show a
transition from disk torque to dipole radiation torque.


\section{SPIN AND LUMINOSITY EVOLUTION OF \418}

We have carried out a detailed investigation of \418 using the
code developed earlier (Ertan \& Erkut 2008; Ertan et al. 2009) which
successfully generated AXP and SGR properties at their likely ages
by luminosity and spindown evolution driven by a fallback disk.
Many combinations of initial conditions were tried in search
for a scenario to produce the present day \418. Each calculation
starts with a choice of dipole moment and initial rotation period
for the neutron star, and an initial disk mass.

The disk around \418 cannot still be inside the light cylinder
at present: if it were, $\Pdot$ would be approximately (or exactly,
in our torque model) independent of $\Mdotin$ in this epoch, so that
the age estimate would be given by $\sim P/\Pdot = 5 \times 10^{7}$
y, an untenably old age. We find that \418 was spun-down to its
period with efficient disk torques in a past epoch when the inner
disk was within the light cylinder, $\Pdot$ having subsequently
decreased to its present value in the present epoch when the disk is
at or beyond $\rlc$.

Figures 1 and 2 show evolutionary tracks for luminosity,
period and period derivative, producing the present properties of
\418 for numerous combinations of the initial conditions.
Figure 1 shows the evolutionary models with $P_0 = 150$
ms for $B_0 = 1.2, 1.4, 1.6 \times 10^{12}$ G, with initial disk
masses $M_d = 5.6, 2.2, 1.3 \times 10^{-5} \Msun$ respectively. In
Figure 2, we show evolutionary tracks with $B_0 = 1.2 \times
10^{12}$ G and $P_0 = 70 - 300$ ms, with $\Md \simeq 6 \times
10^{-5} \Msun$. A reference luminosity $L_x = GM\Mdotin/R$ is
plotted throughout the past epoch when the inner disk was inside
the light cylinder. The true luminosity was less by the unknown
fraction $\Mdotacc/\Mdotin$. This uncertainty does not influence
the evolutionary models because its effect on the disk is folded
into the irradiation efficiency. We took $B_0$ in the
$10^{11}-10^{13}$ G range of dipole fields for most young pulsars,
which worked in earlier applications (Ertan et al. 2009). All AXPs and SGRs
have $L_x \lesssim 10^{36}$ erg/s, giving an upper limit for $M_d$
in our searches. $\Md$ is calculated for disk models extending to
an outer radius $\rout = 5 \times 10^{14}$ cm at the start. For
given $B_0$, disks lighter than a certain $M_d$ can never
penetrate the light cylinder, and so cannot produce an AXP/SGR.
For each $B_0$ - $M_d$ choice, there is a minimum $P_0$ for the
inner disk to ever lie within $\rlc$. The degeneracy of initial
conditions producing \418 shows that these correlations between
workable initial conditions are not very strict constraints. For
most $B_0$ and $M_d$, models start off with the inner disk within
the light cylinder. (Models with stronger $B_0$ and lower $M_d$
show different early evolution, with $\rA > \rlc$ initially.
Starting off under the dipole spin-down torque, the low luminosity
$\sim 10^{34}-10^{35}$ erg s$^{-1}$ in the initial phases is due
to cooling (Page 2009) and dissipative dynamics inside the neutron
star (Alpar 2007). These models show sudden luminosity and torque
increase at $\sim 10^3$ yrs when the inner disk enters the light
cylinder.)

Between $10^4$ and $10^5$ y, there is a turnover to fast
luminosity decay with rapid spin-down until the period settles to
its present value of 9.1 s. The fall-back disk is evolving towards
its final passive phase. As $\Mdotin$ drops, so does the rate of
viscous heating. Effects of irradiation also start to drop as the
accretion luminosity decreases with mass inflow rate. 
Starting from the outermost parts, more and more sections of the
disk are cooling below the critical temperature $\Tp$. As this
continues, $\Mdotin$ arriving at the inner disk to provide for
accretion decreases even more rapidly. The positive feedback leads
to a luminosity turnover and eventual cutoff. Throughout this
phase, $\rA$ is inside the light cylinder and the disk torque
remains in effect. The light cylinder recedes as the star spins
down, but the inner disk recedes more rapidly with the accelerated
decay of $\Mdotin$, and finally reaches $\rlc$. The $\Pdot$ now
starts dropping very rapidly and the period remains almost
constant from this point on. The luminosity is down to the cutoff
luminosity, which we take to be $2 \times 10^{31}$ erg s$^{-1}$,
three times less than the slowly decaying present luminosity
quoted by Rea et al. (2010) for a distance of 2 kpc, and consistent with the
standard cooling luminosity range of neutron stars at ages of
$10^5 - 10^6$ y. The choice of luminosity cutoff does not effect
the evolution.

 The optical and infrared emission of the disk around \418 at present
is much weaker than for other AXPs and SGRs.
We expect luminosities in $K_s$ and 4.5 $\mu m$ bands about $10^3$
and $10^5$ times less than the corresponding luminosities of \142.
The disk luminosity is even lower in the R band, since
the magnetosphere truncates the inner disk of \418.


\section{DISCUSSION AND CONCLUSIONS}

\418 was spun down to its present period in an earlier epoch
when the inner disk was within the light cylinder. The present
state of exceedingly low spindown rate was reached when the disk
retreated to or beyond the light cylinder. Initial parameters
$B_0 \simeq 1 - 2 \times 10^{12}$ G, $\Md \simeq 4 \times 10^{-6}
\Msun$ to $\Md \simeq 6\times 10^{-5} \Msun$ and $P_0 > 70$ ms
work well, giving the period $P_0 = 9.1 $ of \418, consistently
with the upper limit $\dot{P} < 6 \times 10^{-15}$ s s$^{-1}$ at
ages greater than about $2 \times 10^5$ years. For the
present epoch with the disk inner edge at or beyond the light
cylinder tracks with a sustained disk torque are shown, as well as
evolution reduced to dipole spin-down. The luminosity is due to
partial accretion until $t\sim 3 - 6 \times 10^4$ y. For
simplicity we show only a reference luminosity calculated for full
accretion; the actual luminosity in this past epoch was smaller by
an unknown fraction $\Mdotacc/\Mdotin$. The period $P$ = 9.1 s is
reached as an eventual constant period, already at $3 - 6 \times
10^4$ y, together with a drop in period derivative.

Figures 1 and 2 show that the present $\Pdot$ upper limit gives a lower
limit of $\sim 2 \times 10^5$ y if the disk torque is still
operating. If the disk is already out of contact with the star, the
dipole spin-down track gives a lower limit of $\sim  10^5$ y. A
future measurement of $\Pdot$ will give a rough estimate of the age,
between this lower bound and the age at which  the disk torque
models give the observed $\Pdot$. A measurement of $\Pdot \sim 4
\times 10^{-17}$ s s$^{-1}$ will establish dipole spin-down prevails
at present. An even lower $\Pdot$ measurement would signal dipole
spin-down driven by $B_0 < 10^{12}$ G.

We conclude that the very low period derivative upper limit for \418
can be naturally explained in terms of spindown by a fallback disk.
The neutron star has initial rotation period in the range expected
for young neutron stars  (Faucher-Gigu\'ere \& Kaspi 2006). The 
dipole component of the surface field is in the 10$^{12}$ G range. 
The higher multipoles 
and the total surface field could be much larger. Indeed, 
the X-ray spectrum of \418 indicates a total surface 
field of 1.1 $\times$ 10$^{14}$ G (G\"{u}ver, G\"{o}\u{g}\"{u}\c{s} 
\& \"{O}zel 2011).
Comparative investigation of total and
surface dipole magnetic fields by different methods is likely to
provide important clues to properties and evolution of magnetars,
pulsars and young neutron stars.
\acknowledgements
We acknowledge research support from Sabanc\i\ University and from
T\"{U}B{\.I}TAK grant 110T243. M.A.A. thanks the Astronomical
Institute Anton Pannekoek of the University of Amsterdam for
hospitality, the NWO for a grant during his sabbatical in
Amsterdam, and the Turkish Academy of Sciences for research
support. This work was supported by the EC FP6 Marie Curie
Transfer of Knowledge Project ASTRONS, MKTD-CT-2006-042722. We
thank Y. Ek\c{s}i, H. Erkut and E. G\"{o}\u{g}\"{u}\c{s} for
useful discussions.


\begin{figure}
\includegraphics[height=.6\textheight,angle=270]{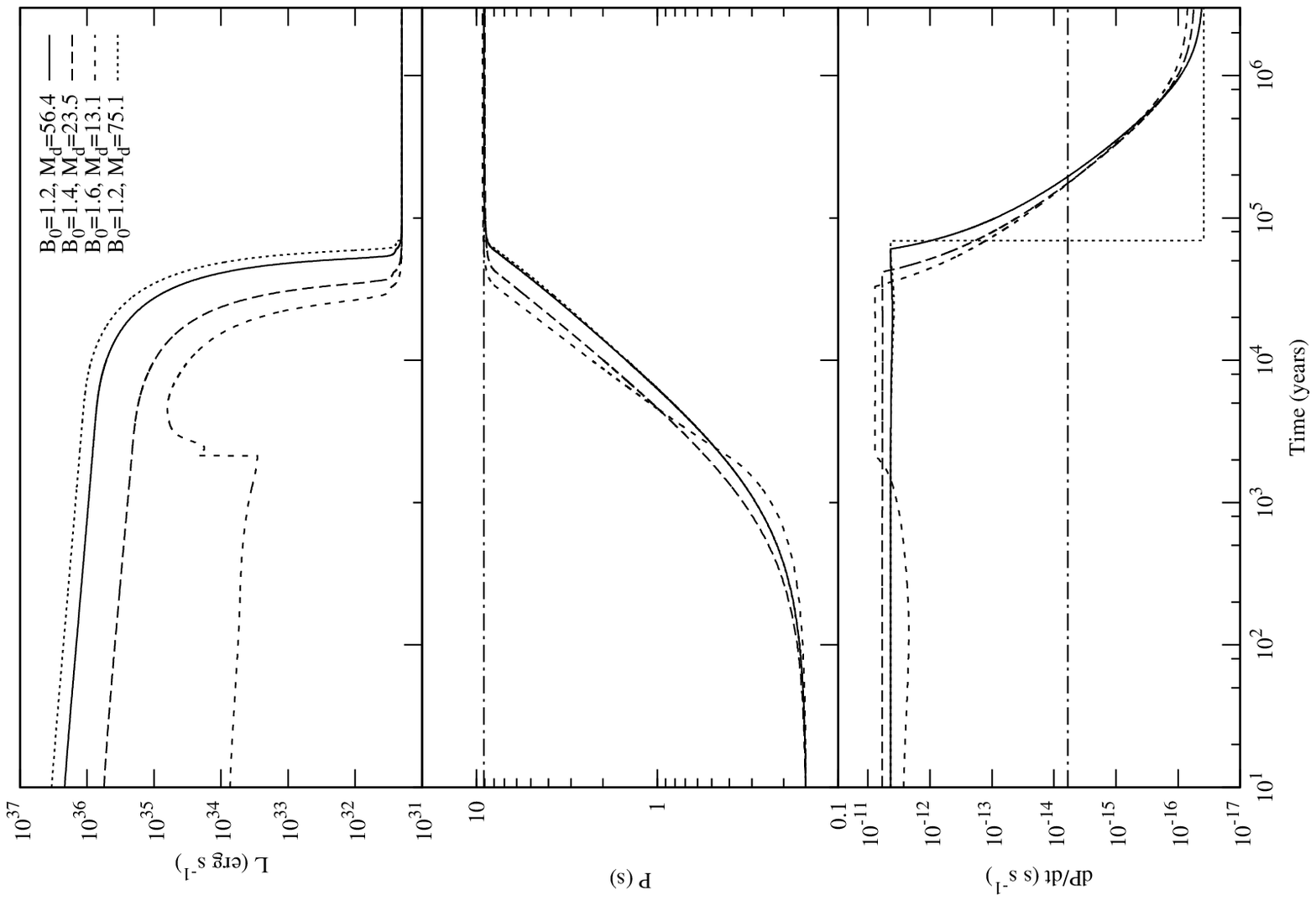}
\caption{Luminosity, period and period-derivative evolution of
model sources for an initial period $P_{0}$ = 150 ms. Values of
the initial disk mass (in units of $10^{-6} \Msun$) and the
magnetic field (in $10^{12}$ Gauss) at the poles of the neutron star are
given in the figure. The horizontal lines correspond to the period
(9.1 s) and the present upper limit on the period-derivative of
\418 ($6 \times 10^{-15}$ s s$^{-1}$). We also present the minimal
torque case (dotted curve)
where the disk torque is assumed to vanish when $r_{A} \geq r_{LC}$.}
\end{figure}

\begin{figure}
\includegraphics[height=.6\textheight,angle=270]{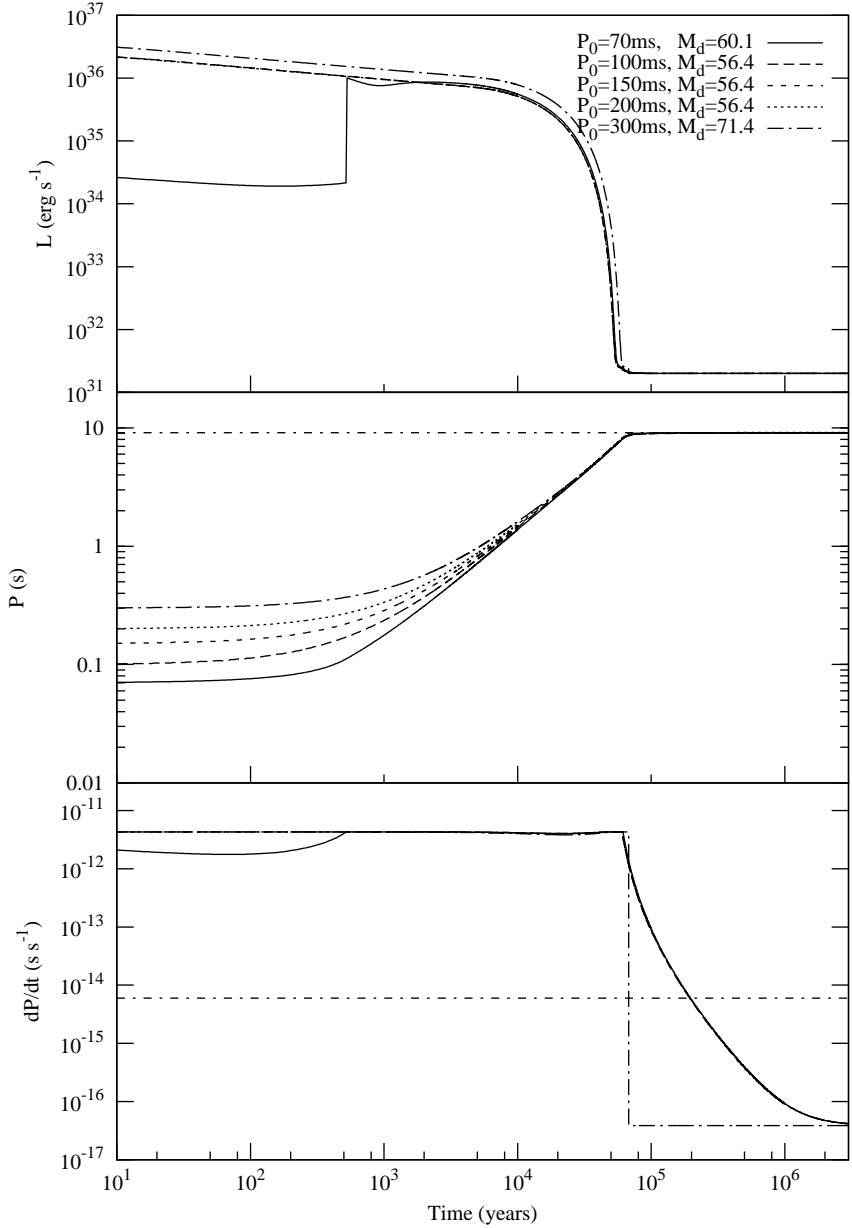}
\caption{Luminosity, period and period-derivative evolution of
model sources for a polar magnetic field of $B_{0}$ = 1.2 $\times
10^{12}$ G on the surface of the neutron star. Values of the
initial disk mass (in units of $10^{-6} \Msun$) and initial period
are given in the figure. The horizontal lines show the present
period and upper-limit on $\Pdot$. The period derivative curves
converge to a final value of $\sim$ 4 $\times 10^{-17}$ s
s$^{-1}$,
a lower limit given by the dipole spin-down torque when
the disk becomes inactive. The dipole spin-down case is given by the
dotted-dashed curve.}
\end{figure}

\end{document}